\documentclass[aps,pre,reprint,groupedaddress]{revtex4-1}

\usepackage{graphicx}

\usepackage{siunitx}

\newcommand{\micron}{\si{\micro\meter}}
\newcommand{\cP}{\mathrm{mPa\cdot s}}

\begin{document}

\title{From splashing to bouncing: the influence of viscosity on the impact of suspension droplets on a solid surface}

\author{Martin H. Klein Schaarsberg$^1$}
\author{Ivo R. Peters$^{1,2}$}
\email[]{i.r.peters@soton.ac.uk}
\author{Menachem Stern$^1$}
\author{Kevin Dodge$^1$}
\author{Wendy W. Zhang$^1$}
\author{Heinrich M. Jaeger$^1$}
\affiliation{
$^1$James Franck Institute, The University of Chicago, Chicago, Illinois 60637, USA\\
$^2$Engineering and the Environment, University of Southampton, Highfield, Southampton SO17 1BJ, UK}

\date{\today}

\begin{abstract}
We experimentally investigated the splashing of dense suspension droplets impacting a solid surface, extending prior work to the regime where the viscosity of the suspending liquid becomes a significant parameter. The overall behavior can be described by a combination of two trends. The first one is that the splashing becomes favored when the kinetic energy of individual particles at the surface of a droplet overcomes the confinement produced by surface tension. This is expressed by a particle-based Weber number $We_p$. The second is that splashing is suppressed by increasing the viscosity of the solvent. This is expressed by the Stokes number $St$, which influences the effective coefficient of restitution of colliding particles. We developed a phase diagram where the splashing onset is delineated as a function of both $We_p$ and $St$. A surprising result occurs at very small Stokes number, where not only splashing is suppressed but also plastic deformation of the droplet. This leads to a situation where droplets can bounce back after impact, an observation we are able to reproduce using discrete particle numerical simulations that take into account viscous interaction between particles and elastic energy.
\end{abstract}

\pacs{}

\maketitle

\section{Introduction}

The study of the impact and splashing of liquid droplets has a long history, from the early work of Worthington over a century ago~\cite{Worthington1908} to recent studies investigating the details of the dynamic spreading and breakup process, including how it is affected by the presence of ambient gases and by modifying the properties of the impacted surface~\cite{Fedorchenko2004,Xu2005,Bolleddula2010a,Driscoll2010,Tsai2010a,Mishra2011,Driscoll2011,Mandre2012,Guemas2012,Latka2012,Bischofberger2013,Josserand2016}. Suspending small, solid particles inside the liquid can fundamentally change the impact dynamics~\cite{Nicolas2005,Luu2009,Qi2011,Guemas2012,Peters2013e,Lubbers2014,Grishaev2015}. In particular, the established criteria for the onset of splashing in pure liquids are no longer valid and until recently it was not even clear whether the presence of particles would increases or decrease the splashing propensity~\cite{Peters2013e}. A major surprise from recent work on concentrated suspensions has been that many factors strongly affecting pure liquid impact, such as substrate microstructure or ambient gas pressure, play only a minor role, if at all~\cite{Peters2013e}. The fact that the impact dynamics of concentrated suspensions is simpler and easier to control than that of pure liquids opens up new possibilities for material deposition applications, including coating and additive manufacturing~\cite{Derby2010}.

Before impact, the suspension droplet is confined purely by surface tension, and the strength of this confinement is tested when the droplet hits a solid surface. Upon impact, this droplet can spread out, or splash through the ejection of particles \cite{Nicolas2005,Peters2013e,Marston2013}, or at the limit of high impact speeds spread out into a monolayer~\cite{Lubbers2014}. Because the impact causes strong deformation on a short time scale, the non-Newtonian behavior of the suspension is expected to play a large role \cite{Guemas2012}.

In a previous publication~\cite{Peters2013e}, we have shown that the global energy budget of the suspension does not influence the onset for the splashing of the suspension droplets, but is instead set by the energy barrier that individual particles need to cross to escape the droplet. More precisely, the ratio between the kinetic energy that particles obtain as a result of collisions and the surface energy associated with an escaping particle needs to be large enough, which can be expressed by a particle-based Weber number $We_p=\rho_pr_pU^2/\sigma$, where $\rho_p$ is the density of the particle, $r_p$ the particle radius, $U$ the impact speed, and $\sigma$ the surface tension of the liquid. In the inviscid limit, the splashing onset was found to be at $We_p\approx14$~\cite{Peters2013e}.

\begin{figure}
    \centering
    \includegraphics{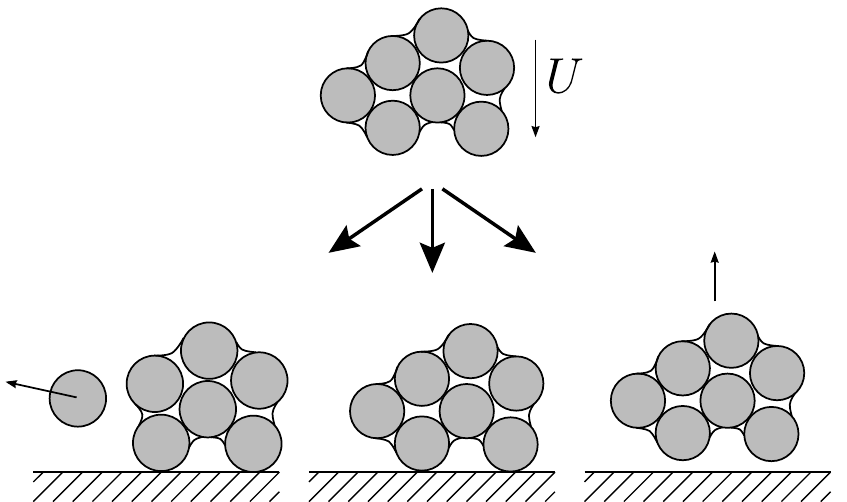}
    \caption{Schematic view of three different droplet impact scenarios: Particle ejection (splashing), spreading without splashing, and bouncing.}
    \label{fig:impactScenarios}
\end{figure}

Here, we explore with experiments and supporting simulations what happens if we increase the dissipation in this system by changing the viscosity of the suspending liquid. Previous studies \cite{Gondret2002,Kantak2004,Yang2006a,Donahue2010a,Gollwitzer2012} have shown that the effective coefficient of restitution $e$ is a function of the Stokes number $St=\frac{2}{9}\rho_p r_p U/\mu$, with $\mu$ the dynamic viscosity of the liquid. A common feature is that $e$ drops to zero at a critical value of the Stokes number $St_c$. In fully submerged systems, particle impact on a flat surface in an ambient fluid~\cite{Gondret2002} and particle-particle collision in an ambient fluid~\cite{Yang2006a} result in $St_c\approx10$. For the case of thin liquid films the results are more involved: typically the critical Stokes number is lower, e.g. $St_c\sim1$ was reported in the case of the impact of dry particles on a wetted surface, although it was shown later by Gollwitzer \emph{et al.}~\cite{Gollwitzer2012} that the critical Stokes number depends on the ratio between the film thickness $\delta$ and the particle radius, where $St_c$ decreased with the ratio $\delta/r_p$. We will use the general notion of the dependence of $e$ on the Stokes number to rationalize the influence of viscosity on the splashing onset of dense suspension droplets. Assuming that splashing is caused by collisions between particles, decreasing $e$ should increase the onset impact speed for splashing. In addition, a critical Stokes number is expected below which splashing is suppressed completely.

\begin{figure}
	\centering
	\includegraphics{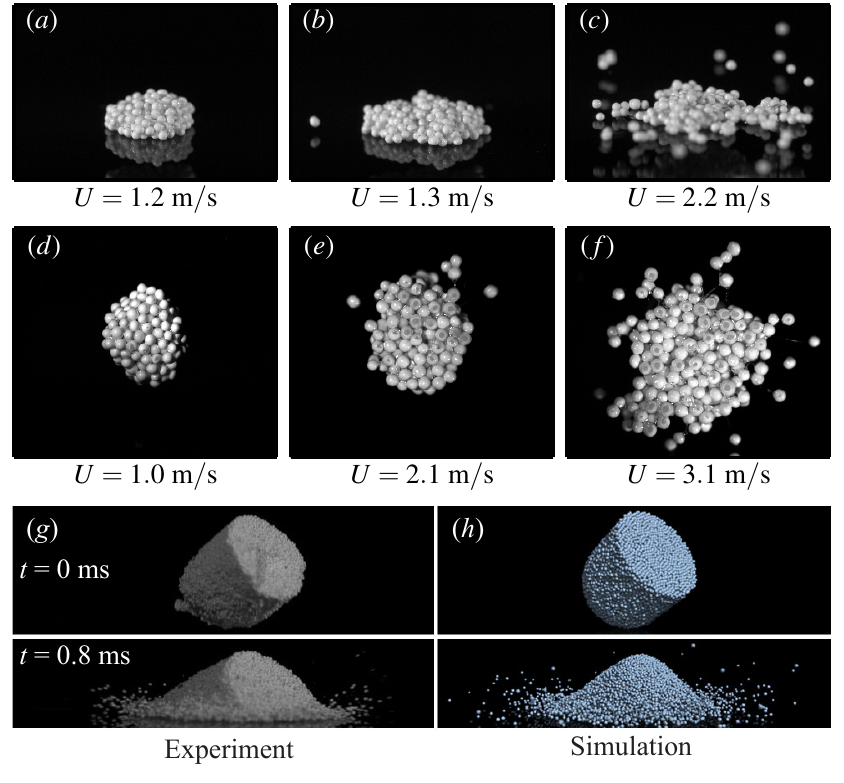}
	\caption{Snapshots of suspension droplets impacting on a glass substrate. ($a$-$c$): side views, $\mu=19~\mathrm{mPa\cdot s}$, $r_p=362~\mathrm{\si{\micro}m}$. ($d$-$f$): bottom views, $\mu=91~\mathrm{mPa\cdot s}$, $r_p=362~\mathrm{\si{\micro}m}$. ($g$, $h$): Comparison between experiment and three-dimensional simulation at $We_p\rightarrow\infty$ and $St\rightarrow\infty$. Panel ($g$) adapted from \cite{Lubbers2014}.}
	\label{fig:snapshots}
\end{figure}

There are three possible scenarios for a suspension droplet impacting a hard surface with speed $U$, as shown schematically in Fig.~\ref{fig:impactScenarios}: it can splash by ejecting particles or, without splashing, it can either stick or bounce back. Figure~\ref{fig:snapshots} shows examples of this general set of behaviors. As the impact speed increases, a splashing threshold is reached beyond which individual particles become ejected. Comparison of the top two rows of images demonstrates that the threshold speed increases in a nontrivial manner with solvent viscosity. At sufficiently large impact velocity, any particle confinement due to surface tension becomes negligible compared to inertia and frictional particle-particle interactions. In this case the behavior is well approximated by the limit of infinite particle-based Weber number. This is demonstrated by the bottom row in Fig.~\ref{fig:snapshots}, which compares side-by-side experiment and results from a simulation of a granular droplet without any liquid included.

Our main experimental result is the behavior of the splashing threshold for increasing solvent viscosity, for which we determined the onset values of the Stokes and particle-based Weber numbers. In addition we have found a regime where the suspension droplets neither splash nor spread, but instead bounce back. The paper is organized as follows. We start by describing the experimental methods in section \ref{sec:experiments} and the numerical model in section \ref{sec:numerical}. After that, we describe in section \ref{sec:results} the experimental observations for splashing and bouncing, and compare the bouncing behavior of viscous suspensions with numerical simulations. We end with a discussion in section \ref{sec:discussion}.

\section{Experiments}\label{sec:experiments}

\begin{figure}
	\centering
	\includegraphics{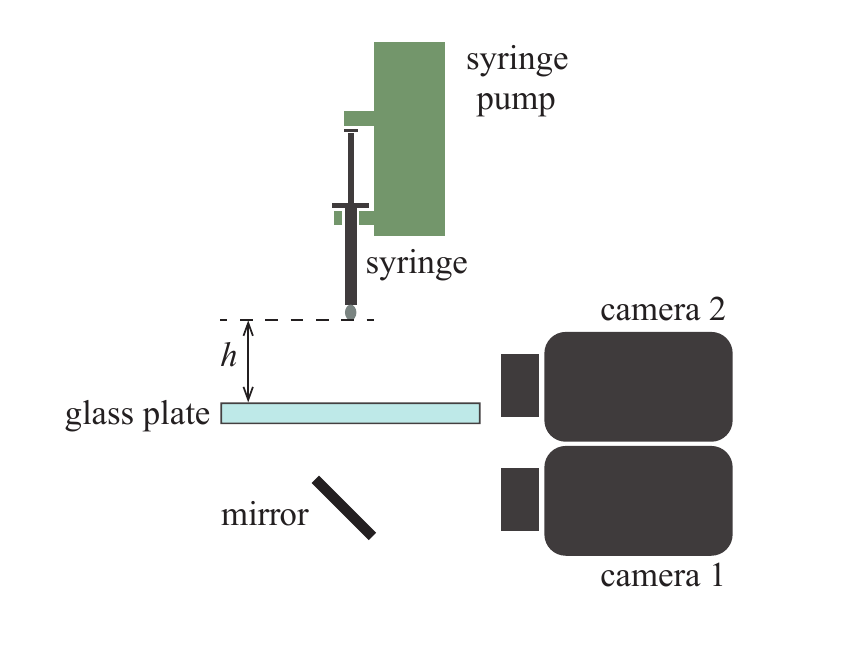}
	\caption{Schematic view of the setup. Suspension droplets impact the glass plate at a speed $U$, set by the release height $h$. The ejection of particles and bouncing motion of the droplets upon impact is recorded using two high-speed cameras.}
	\label{fig:setup}
\end{figure}

We performed experiments where we impacted suspension droplets on a glass plate, and observed their splashing behavior using high speed imaging (Fig.~\ref{fig:setup}). The main control parameters are the impact speed $U$, particle size $r_p$, particle density $\rho_p$, dynamic liquid viscosity $\mu$, and liquid surface tension $\sigma$. Below we explain in more detail how we varied each of these parameters. The parameters can be combined into two relevant dimensionless numbers that we defined in the introduction, the particle-based Weber number $We_p$ and the Stokes number $St$.

The suspensions were created by first filling a syringe with a solvent liquid, followed by carefully adding particles and letting them sediment, while taking care that no air bubbles were entrained. Particles were added until the liquid was fully saturated with particles, which resulted in an average volume packing fraction for all our experiments of $\phi=0.59\pm0.04$. For suspensions made with glycerol, we determined the packing fraction of the resulting suspension droplets by evaporating the liquid on a hot plate and measuring the mass before and after evaporation. We used zirconium dioxide (ZrO$_2$) and soda-lime glass beads for our suspended particles. The properties of these particles, which are a subset of the particles used in \cite{Peters2013e}, are listed in Table~\ref{tab:particles}.
\begin{table}[b]
\caption{\label{tab:particles}
Properties of the particles used in this study
}
\begin{ruledtabular}
\begin{tabular}{crrr}
\textrm{Particle}&
\textrm{Material}&
\textrm{Radius ($\si{\micro\meter}$)}&
\textrm{Density ($\mathrm{kg/m^3}$)}\\
\colrule
$P_1$ & Glass & $76\pm13$ & $2520\pm160$\\
$P_2$ & ZrO$_2$ & $138\pm11$ & $3840\pm160$\\
$P_3$ & ZrO$_2$ & $362\pm22$ & $3930\pm160$\\
\end{tabular}
\end{ruledtabular}
\end{table}

In order to get a range in both viscosity and surface tension, we used silicone oils and glycerol/water mixtures. The silicone oils had viscosities in the range of 20 to $1055~\cP$ and a surface tension around $21~\mathrm{mN/m}$. Glycerol/water mixtures had viscosities ranging from 1 to $1090~\cP$, and surface tensions from $72~\mathrm{mN/m}$ for pure water, decreasing down to $64~\mathrm{mN/m}$ for the most viscous mixtures (99 wt\% glycerol).

The impact velocity was controlled by varying the release height $h$ from 1 to $180~\mathrm{cm}$, giving impact velocities between 0.44 and $5.9~\mathrm{m/s}$. The droplets were created by extruding them quasi-statically from their syringe (inner diameter 4.7 mm) using a syringe pump (Razel R99-EB) at a flow rate of the order of $\SI{1}{\micro\liter/\second}$. The syringes had their tips cut off such that the suspensions were extruded from a simple straight cylinder and did not need to flow through a contracting nozzle. During extrusion, the droplets pinched off under the influence of their own weight, which resulted in reproducible drop volumes. The bottom of the drops have a cylindrical shape with a diameter equal to the syringe inner diameter because they do not deform significantly during the extrusion, while the top of the drops have a sharp tip as a result of the pinch-off process~\cite{Miskin2012,Peters2013e}.

The experiments were recorded with a Phantom V12 high-speed camera operating at a frame rate of 6200 frames/sec (camera 1 in Fig.~\ref{fig:setup}). Using a 105 mm Nikkon Micro-Nikkor lens, the typical resolution of our images was $\SI{20}{\micro\meter}\mathrm{/pixel}$. All experiments were recorded with a bottom view, which was for part of the experiments supplemented by a synchronized side view using a second high-speed camera (camera 2 in Fig.~\ref{fig:setup}). Bottom views were used to determine the splashing onset, as these images are the most reliable for detecting ejected particles. Side view images, on the other hand, allowed us to observe bouncing motion of droplets.


\section{Numerical model}\label{sec:numerical}
In the simulation, we idealize the dense suspension as a collection of elastic spheres experiencing viscous drag due to lubrication flow in the narrow gaps between the particles, surface tension, and particle inertia. Assuming that physical contact between particles occurs via surface asperities characterized by roughness lengthscale $R_\ell$, the lubrication drag between two particles in relative motion has the form $ {\bf F}_\mu=(3\pi/2)({\mu\Delta U r_p^2}/(s + R_\ell)) {\bf n}$~\cite{Ball1995}. Here $\Delta U$ is the relative velocity along the line of approach in the particle pair's center of mass frame, $s$ the separation between particle surfaces, and $\bf n$ the direction along the line of approach. Both $\mu$ and $r_p$ values are chosen in accordance with the experimental values.  We use $R_\ell = 10^{-2}r_p$, corresponding to $\si{\micro}$m-scale roughness over $100$ $\si{\micro}$m-sized particles. When particles collide so energetically that they support nonzero overlap $\delta$, we expect that the lubrication drag saturates at $(3\pi/2) \left( {\mu \Delta U r_p^2}/{R_\ell} \right)$, the value corresponding to a narrow gap at the surface roughness lengthscale.  Particles in the suspension plug are not elastically compressed initially but are compressed by the impact process. We assume that the elastic compression $\delta$ remains small relative to the particle radius $r_p$, therefore giving rise to a Hertzian contact force ${\bf F}_\delta = (2E/3) \sqrt{2r_p} \delta^{3/2} {\bf n}$ between neighboring compressed particles. It was not practical to simulate the impact for the $250$ GPa Young's modulus value associated with ZrO$_2$ particles. Instead we use $E=30$ to $100$ MPa. We include surface tension effects as a bridging force ${\bf F}_\sigma$ between neighboring particles on the surface of the suspension plug. Initially, all surface particles are densely packed together and experience a nonzero capillary bridging force. As impact proceeds and distances between particles grow, this capillary bridging force vanishes once the separation $s$ exceeds a critical value $s_c$.  The simulation uses ${\bf F}_\sigma = 2 \pi \sigma r_p \alpha / (1 + c_1 \hat{s} + c_2 \hat{s}^2 ) \bf n$, where $\sigma\approx70~\mathrm{mN/m}$ is the surface tension of the air-water surface, $c_1 \approx 1.05$ and $c_2 \approx 2.5$ and $\hat{s} = {s}/{[2r_p\sqrt{\pi ( \sqrt{3}/2 - 5/6)}]}$. This expression has the phenomenological form proposed by Herminghaus \emph{et al.}~\cite{Herminghaus2005} to describe a static, axisymmetric capillary bridge between two particles. Since impact occurs at large particle-based Weber number, where particle inertia is important, we expect $\alpha$ and $s_c$ values to differ considerably from the static limit values.  We therefore picked $\alpha$ and $s_c$ values to reproduce the splashing onset at high $St$ where we recover the onset at $We_p\approx14$ for one set of experimental observations ($\alpha = 2.9$ and $s_c = 0.3r_p$).  The parameter values are then left fixed for other $We_p$ values. In addition, implementing ${\bf F}_\sigma$ requires that we accurately flag the surface particles. We do so via a surface detection scheme that first creates a smoothed density field, then calculates the gradient of the density at the center of each particle.  When the density gradient exceeds a cut-off value, the particle is flagged as a surface particle.

To simulate an impact, we start with a collection of elastic spheres in random close packing and prescribe uniform downward speed $U_0$ for all the particles. Upon collision with a rigid wall, we assume that the surface particles are weakly interacting with the wall (asides from facilitating elastic compression). We model this assumption with a modified effective viscosity at the wall $\beta\mu$, where $\beta$ is set to a value between 0 and 1. The weak lubrication interaction with the wall is then characterized by a wall-based Stokes number $St_w=\frac{2}{9}\rho_p r_p U/(\beta\mu)$, and the ratio $St_w/St$ controls the ratio of lubrication interaction strengths between the bulk and the wall. As the impact proceeds, the suspension may flatten and expands radially creating new surfaces.

The three-dimensional simulations in Fig.~\ref{fig:snapshots} were performed for the infinite particle-based Weber number limit, where the presumable role of the interstitial liquid, including confinement due to surface tension, is no longer significant. In this limit, the drop is simulated as an aggregate of dry grains experiencing inelastic, frictionless collisions~\cite{Guttenberg2012,Ellowitz2013} which are characterised by a constant coefficient of restitution of 0.95.


\section{Results}\label{sec:results}

\subsection{Splashing onset}
\begin{figure}
	\centering
	\includegraphics{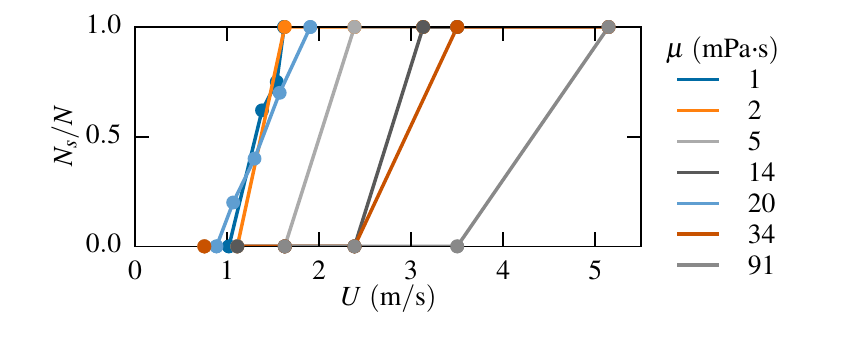}
	\caption{The splashing onsets for $r_p=138~\mathrm{\mu m}$ ZrO$_2$ particles at different liquid viscosities. All liquids are water/glycerol mixtures, ranging from 0 to 85 wt\% glycerol, except the $20~\cP$ solution (light blue), which is a silicone oil. Data for $\mu=1~\mathrm{mPa\cdot s}$ from \cite{Peters2013e}.}
	\label{fig:splashOnsetSpeeds}
\end{figure}

Similar to \cite{Peters2013e}, we defined a splashing event as the ejection of one or more particles from the suspension droplet. We determined the splashing threshold for a specific particle/liquid combination by increasing the impact speed step by step, and repeating each impact speed, depending on the experiment, 3 to 10 times. For each impact speed, we counted the number of observed splashes $N_S$ and divide by the number of repeats of the experiment $N$, with which we define the splashing probability $N_S/N$. Figure~\ref{fig:splashOnsetSpeeds} shows the splashing onsets for the $138~\micron$ ZrO$_2$ particles ($P_2$) for different suspending liquids. We only included data where we were able to find the complete transition from $N_S/N=0$ to $N_S/N=1$.

Clearly, the splashing onset velocity increases with increasing viscosity, although there is no significant difference between the $1~\cP$ and $2~\cP$ solvent viscosity. The only exception is the silicone oil (viscosity $20~\cP$), which has a splashing onset velocity comparable to that of water, and significantly lower than the $5~\cP$ water/glycerol mixture. The reason for this is the lower surface tension of the silicone oil (approximately a factor 3 difference), which decreases the splashing onset velocity.

\begin{figure}
	\centering
	\includegraphics{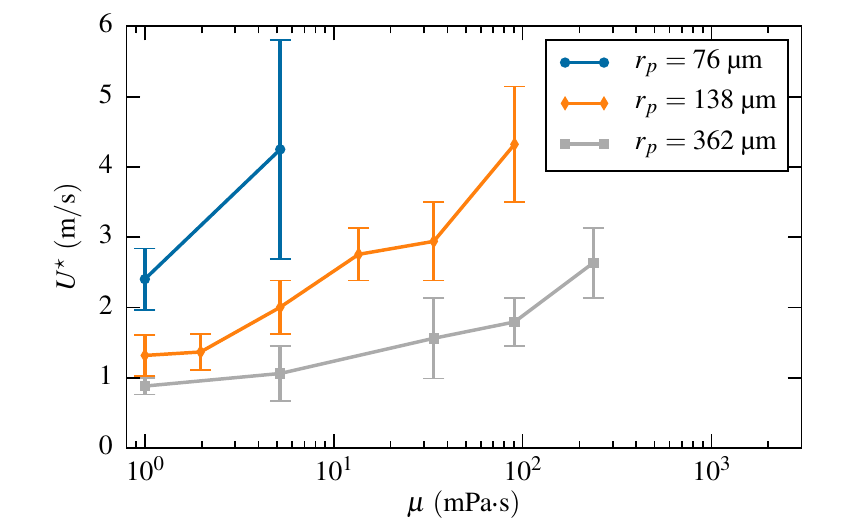}
	\caption{Splashing onset speed $U^*$ as a function of viscosity $\mu$, for the two different sizes of ZrO$_2$ particles (orange, grey) and the glass particles (blue). All data are for water/glycerol mixtures, giving only small variations in surface tension. Data for $\mu=1~\mathrm{mPa\cdot s}$ from \cite{Peters2013e}.}
	\label{fig:onsetSpeedViscosity}
\end{figure}

Each line in Fig.~\ref{fig:splashOnsetSpeeds} represents a splashing onset velocity $U^*$ with an associated uncertainty. The orange data in Fig.~\ref{fig:onsetSpeedViscosity} corresponds to the data in Fig.~\ref{fig:splashOnsetSpeeds}, showing the dependence of the splash onset speed on the viscosity of the suspending liquid. We have added data for the larger particles (grey) and the smaller glass particles (blue), which demonstrates that particles with more mass are less sensitive to the increase in viscosity. We have not included data with silicone oils in Fig.~\ref{fig:onsetSpeedViscosity} to show only the influence of viscosity, particle size and particle density.

\begin{figure}
	\centering
	\includegraphics{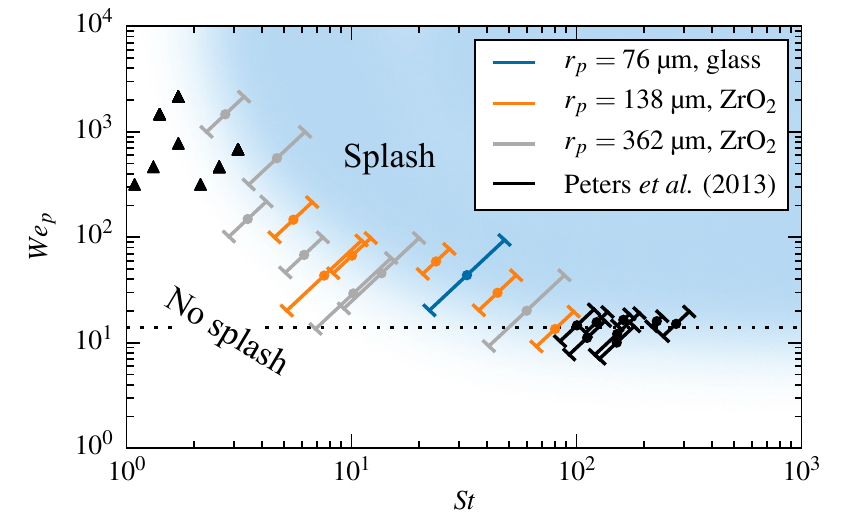}
	\caption{Splashing state diagram with the particle-based Weber number and the Stokes number. Data delineate the transition. Error bars give the width of the transition region (see main text). The horizontal dotted line represents the asymptotic value $We_p\approx14$ for $St\rightarrow\infty$ from \cite{Peters2013e}. The experiments where we have observed bouncing motion are indicated with black triangles. Note that the bouncing data extends to $St<1.0$, and the full range is plotted in Fig. \ref{fig:bouncePlot}.}
	\label{fig:masterCurve}
\end{figure}

We can capture both the influence of surface tension as well as the influence of viscosity by plotting $We_p$ \cite{Peters2013e} versus $St$. All our data for the splashing onset in Fig.~\ref{fig:masterCurve}, which includes glycerol/water mixtures and silicone oils, falls roughly on a single master curve, showing that indeed $We_p$ and $St$ are the relevant parameters determining the splashing onset for our suspensions. The black data in this figure correspond to the data from Peters~\emph{et al.} \cite{Peters2013e} which all are suspensions in demineralized water, with 76, 107, 175, 249, and 359 $\mathrm{\si{\micro}m}$ glass, and 78, 138, and 362 $\mathrm{\si{\micro}m}$ ZrO$_2$ particles. Note that we have determined the splashing onset for each particle/liquid combination by determining the maximum speed at which we never observe a splash and the minimum speed at which we always observe a splash. This results in diagonal error bars, because varying the impact velocity changes both $We_p$ and $St$.

Figure \ref{fig:masterCurve} shows a limiting case represented by the horizontal dashed line, for $St\rightarrow\infty$, where the splashing onset becomes independent of the Stokes number and can be described solely by the threshold value $We_p\approx14$. The reason for this limit is that for high enough Stokes number ($St\gtrsim100$), changes in the effective coefficient of restitution due to viscous effects become small enough to approximate it as a constant value~\cite{Gondret2002}. We speculate that there exists a second limit for $We_p\rightarrow\infty$, where the splashing can be suppressed by viscous effects alone, without the need of surface tension, although our current experimental results are not conclusive on this point.

\subsection{Bouncing motion}
For suspensions droplets with high liquid viscosities we observed bouncing motion in both the experiments and the simulations. This is a surprising result, because increasing the viscosity of the liquid increases the dissipation, while a rebound requires the storage of energy through deformation. Comparing our findings to the rebound of pure liquid droplets, we find a clear difference in the Weber number regime where rebounds are observed. In pure liquid droplets, rebounds are typically observed for intermediate Weber numbers ($0.2 \leq We \leq 60$) \cite{Rioboo2008, Tsai2009, Tsai2011}. Upon impact, the droplets need to have enough inertia to appreciably deform the droplet, but surface tension still needs to be strong enough to prevent the droplet from breaking up and bring it back toward a spherical shape.

The bounces we observed in the suspension droplets all occurred at very high (particle-based) Weber numbers, $100\leq We_p\leq2000$~\footnote{For comparison we may also look at the droplet-based Weber number, but these values will be even bigger as the droplet size is always much bigger than the particle size.}. We also only see very little deformation of the droplets, which suggests that most of the kinetic energy at impact is dissipated due to the viscous liquid and surface tension does not play an important role in the rebound. We can quantify this by estimating the total stored energy needed for the rebound, and calculating the surface area associated with a surface energy that equals the stored energy.

\begin{figure}
	\centering
	\includegraphics{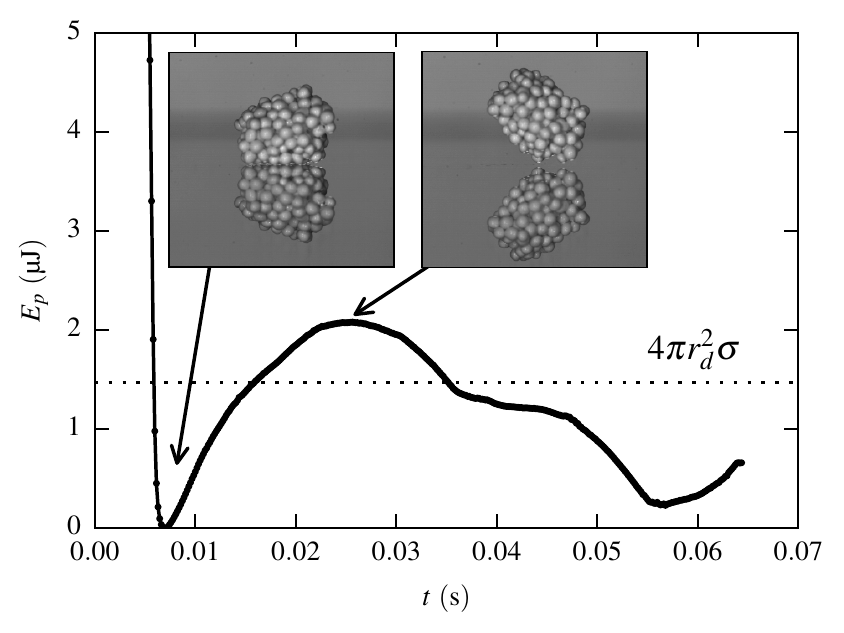}
	\caption{Estimated potential energy of a bouncing droplet ($362~\micron$ ZrO$_2$ in $1055~\cP$ silicone oil). For reference, the horizontal dotted line gives the theoretical surface energy of a silicone oil sphere with radius $r_d$.}
	\label{fig:bouncingEnergy}
\end{figure}
Fig.~\ref{fig:bouncingEnergy} shows the potential energy of a droplet during a bounce. We estimate the change in potential energy by determining the two-dimensional center of mass in the high speed images and get the height $h$ as a function of time, which with the known mass gives the potential energy $E_p\approx4/3\pi r_d^3[\phi\rho_p+(1-\phi)\rho_l]gh$, with $r_d$ the droplet radius, $\rho_l$ the liquid density and $g$ the gravitational acceleration. For the specific case shown here, we found a maximum potential energy of about $2~\mathrm{\si{\micro}J}$. In order to understand which mechanism is able to store this amount of energy during the impact, we compare this to a change in surface area that would store an equivalent amount of surface energy, like the mechanism that is responsible for bouncing pure liquid droplets. The dotted line in Fig.~\ref{fig:bouncingEnergy} gives the surface energy of a spherical droplet of the solvent used in this suspension. This shows that in order to store enough energy for the observed bounce, an excess of surface area needs to be created during the impact of the order of the area of the droplet itself. This is very unlikely since no appreciable deformation of the suspension droplets can be seen in the high speed images (see inset of Fig.~\ref{fig:bouncingEnergy}). In addition, such large deformations would be largely dissipated by the viscosity of the liquid. We therefore expect the energy to be stored in elastic deformation of the particles in the suspension.

\begin{figure}
	\centering
	\includegraphics{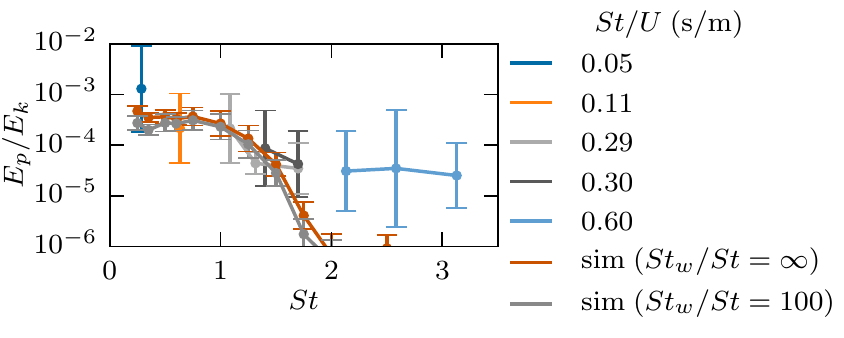}
	\caption{Ratio of stored elastic energy and kinetic energy as a function of the Stokes number, for different viscosities and particle sizes. The error bars represent the standard deviation of repeated experiments, with 3 to 11 repetitions per data point. Two sets of simulation data are shown, for two different values of wall drag.}
	\label{fig:bouncePlot}
\end{figure}

The analysis above was done for a droplet which during the bounce completely detaches from the surface. If we take a less strict approach and define a bouncing drop by any upward motion after the impact, we get a better view of how the viscosity is influencing the probability to find a bouncing droplet. This is shown in Fig.~\ref{fig:bouncePlot}, where we plot the ratio of the kinetic energy over the maximum potential energy ($E_p/E_k=(gh)/(U^2/2)$) as a function of the Stokes number. To show which experiments were performed with the same particle/liquid combination, we color-coded and identified them with a number that is independent of impact speed, \emph{i.e.}, $St/U$. Although the fluctuations in experimental outcomes are large, as the error bars indicate, there is a significant increase in restored energy towards lower values of the Stokes number.

\begin{figure*}
	\centering
	\includegraphics{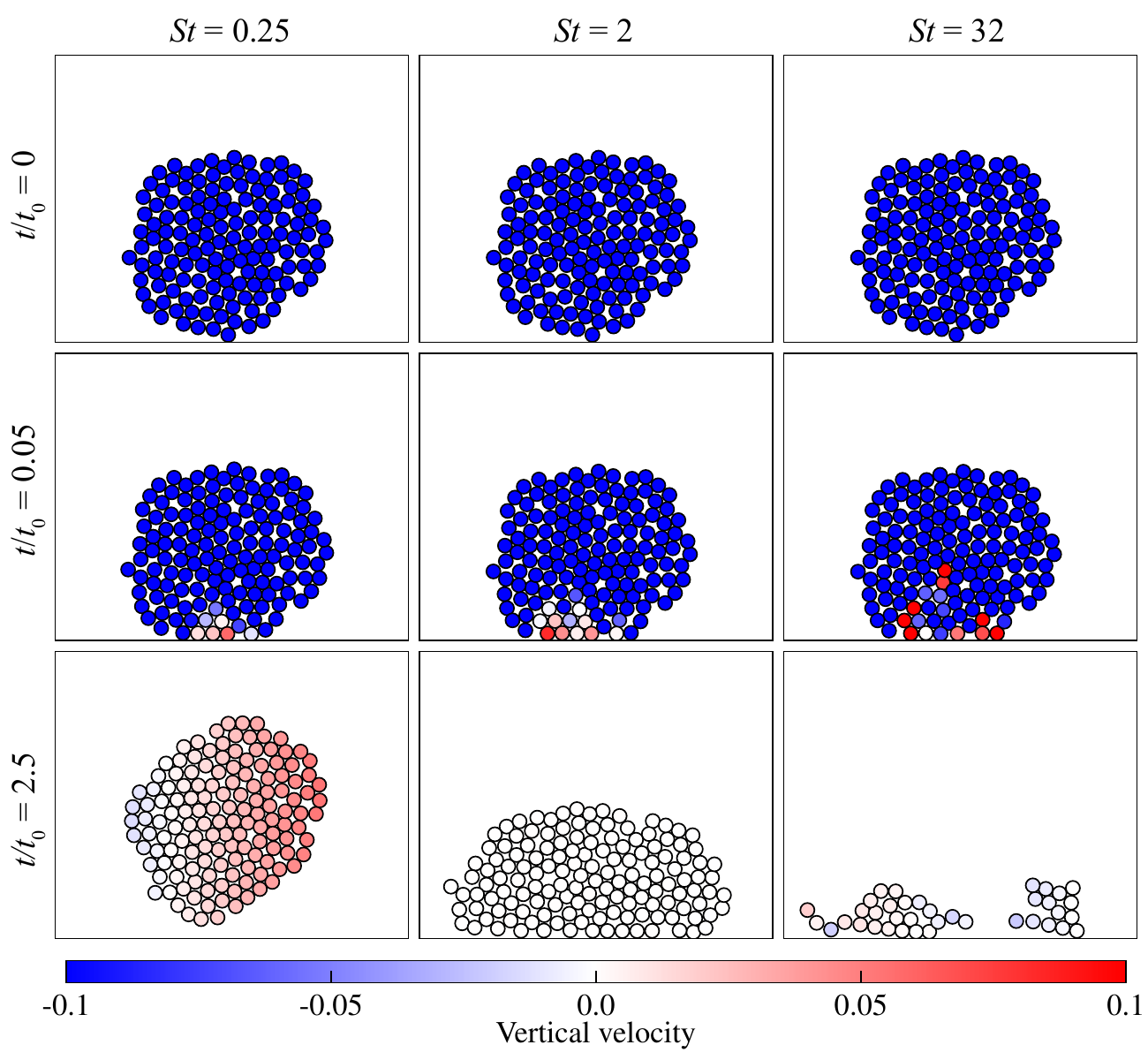}
	\caption{Snapshots from the numerical simulation of two-dimensional suspension droplets at different Stokes numbers (all at infinite Weber number; only the suspended particles are rendered, the liquid is not shown). Time after impact increases from top to bottom with $t=0$ the moment of impact. Colors indicate the vertical velocity component of individual particles. $St=0.25$ shows a bounce, along with a significant amount of rotation.}
	\label{fig:elasticEnergySnapshots}
\end{figure*}
A possible explanation for the observations above would be that the bouncing results from elastic energy that is stored in a jammed network of particles that are in contact. In the case of high viscosity these networks might be stabilized, while at lower viscosity this network would fall apart like it would for the impact of a dry granular droplet. To test this idea we performed numerical simulations of the impact of a two-dimensional suspension droplet. Fig.~\ref{fig:elasticEnergySnapshots} shows snapshots of the numerical simulation for three different values of the Stokes number. In all three cases the Weber number is infinite, such that we can neglect any influence of surface tension. Time and velocity are made dimensionless with the typical time $t_0=r_p/U_0$ and impact speed $U_0$, respectively. Note that the color scale is clipped at $0.1U_0$, to emphasize the velocities after impact. At the lowest Stokes number $St=0.25$ we observe a clear bounce at $t/t_0=2.5$. This in contrast with $St=2$, where all motion is dissipated after impact. At even higher Stokes numbers, the droplet is fragmentized.

In addition to the bounce at low Stokes numbers, we also observe a fair amount of rotation in the droplet (Fig.~\ref{fig:elasticEnergySnapshots}c). This rotation, as expected, depends on the initial orientation of the droplet on the approach to the substrate. This rotation adds to the total restored energy, but was not taken into account in the experimental data (Figs. \ref{fig:bouncingEnergy} \& \ref{fig:bouncePlot}) because we had no accurate method to determine this contribution. Therefore, to compare our simulation results to the experimental data, we calculated the restored energy using the vertical velocity of the center of mass of the suspension droplet after impact. In Fig.~\ref{fig:bouncePlot} we compare the restored energy as calculated in the simulations to the experimental results and find good agreement for $St\lesssim1.5$.


\section{Discussion \& conclusions}\label{sec:discussion}

Using a wide range of liquid viscosities we experimentally determined the influence of the Stokes number on the splashing onset of dense suspensions. We see a significant departure from the inviscid limit at $St\lesssim50$, where the particle-based Weber number at the splashing onset becomes larger than 14 within error bars. At the lowest Stokes numbers where we were able to observe a splashing onset, we find an increase of about two orders of magnitude of $We_p$ compared to the inviscid limit. The steep increase of the splashing onset is suggestive of a diverging behavior, although our data are currently not conclusive at this point.

A striking effect for very small values of the Stokes number is the bouncing of the suspension droplets. Because a higher viscosity is typically associated with an increase of dissipation, it is surprising that suspensions with a higher solvent viscosity are more likely to bounce than their inviscid counterparts. We quantified this behavior and observed a clear trend of increased stored energy relative to the initial energy with decreasing Stokes number. Our simulations show the same trend, although quantitatively there is a difference in the Stokes numbers at which the bouncing motion becomes significant. A possible explanation for the dependence of the bouncing on the Stokes number could be found by comparing the pressure induced by the impact to the pressure needed to generate a substantial flow of liquid through the densely packed suspension. In Appendix \ref{app:porousFlow} we outline such an argument by treating the suspension as a porous media and using the Kozeny-Carman relation to estimate the flow resulting from a pressure gradient.

Finally, we add a note about cavitation: The separation of the particles from the bottom substrate during a bouncing event will involve a significant drop in pressure due to the lubricating flow. Especially because this happens at high viscosities, this may result in cavitation~\cite{Joseph1998,Marston2011}. Although we cannot exclude that cavitation bubbles are formed during rebound events, we have not observed any signature of cavitation. This would be an interesting path for further research, as cavitation could possibly affect the bounce strength. Cavitation might also be a source for the reduced lubrication interactions with the impacted wall.

\begin{acknowledgments}
We thank Christophe Clanet, Devaraj van der Meer, Tom Witten, and Qin Xu for insightful discussions. M.K.S. thanks the James Franck Institute for hospitality during his stay in Chicago and the Physics of Fluids group for financial support. This work was supported by the Chicago MRSEC, which is funded by NSF through grant DMR-1420709.
\end{acknowledgments}

\appendix
\section{Stokes number dependence for bouncing onset}\label{app:porousFlow}
Here we provide a possible argument for the Stokes number dependence on the bouncing we observe in Fig.~\ref{fig:bouncePlot}. We treat the suspension as a porous media such that we can use the Kozeny-Carman relation for spherical particles with packing fraction $\phi$ to relate the liquid flow speed $u_l$ to a pressure difference $\Delta P$ over a length scale $L$:
\begin{equation}
    \frac{\Delta P}{L} = \frac{45 \mu \phi^2 u_l}{r_p^2 (1-\phi)^3}.
\end{equation}
In our droplet impact problem, we may estimate the pressure from the momentum of a droplet $r_d$ $mU \approx \frac{4}{3}\pi r_d^3\rho U$ (assuming a spherical droplet with radius $r_d$ and average density $\rho$), the impact time scale $\tau$, and the area $A\sim \pi r_d^2$. Approximating the length scale $L$ as $r_d$, and neglecting numerical values of $\mathcal{O}(1)$ gives
\begin{equation}
    \frac{\Delta P}{L} \sim \rho\frac{U}{\tau}.
\end{equation}
We expect the granular packing to lose its stability if the interior liquid is displaced over a typical distance comparable to the particle size $r_p$ during the short impact time $\tau$. This sets the velocity $u_l\sim r_p/\tau$. The condition for bouncing would be if the pressure due to the impact is too small to generate a velocity $u_l$. This gives
\begin{equation}
    \rho\frac{U}{\tau} \lesssim \frac{45 \mu \phi^2}{r_p (1-\phi)^3\tau},
\end{equation}
which can be rewritten as
\begin{equation}
    \label{eq:StokesBounce}
    St=\frac{2}{9}\frac{\rho r_p U}{\mu} \lesssim \frac{10 \phi^2}{(1-\phi)^3},
\end{equation}
and predicts a dependence purely on the Stokes number. Note that the density in (\ref{eq:StokesBounce}) is the average density of the droplet and not the density of the individual particles, which would introduce a correction of $\mathcal{O}(1)$. Using $\phi\approx0.6$ in (\ref{eq:StokesBounce}) predicts an upper limit $St\sim56$, which is clearly much larger than what we observe experimentally. A more detailed analysis of the flow and stability of the granular packing might resolve this, but is outside the scope of the current study.

\end{document}